\documentclass[fleqn,10pt]{wlscirep}
\usepackage[utf8]{inputenc}
\usepackage[T1]{fontenc}
\title{Sub-second Temporal Magnetic Field Microscopy Using Quantum Defects in Diamond}
\newcommand{\etal}{\textit{et al}.\@ }
\newcommand{\ie}{\textit{i}.\textit{e}.}

\usepackage{fancyvrb}
\usepackage[version=4]{mhchem}
\usepackage{hyperref}
\hypersetup{
    colorlinks=true,
    linkcolor=blue,
    filecolor=magenta,      
    urlcolor=blue,
    pdftitle={Sub-second Temporal Magnetic Field Microscopy Using Quantum Defects in Diamond},
    pdfauthor={Madhur Parashar, Anjuj Bathla, Sharba Bandyopadhyay, Kasturi Saha}     
    }
\usepackage{siunitx}
\usepackage{txfonts}
\usepackage{float}
\usepackage{physics}
\usepackage{xcolor}
\author[1,2]{Madhur Parashar}
\author[3,4]{Anuj Bathla}
\author[3]{Dasika Shishir}
\author[3]{Alok Gokhale}
\author[1]{Sharba Bandyopadhyay}
\author[3,*]{Kasturi Saha}
\affil[1]{Information Processing Laboratory, Department of Electronics and Electrical Communication Engineering, Indian Institute of Technology Kharagpur, Kharagpur, West Bengal, India -721302}
\affil[2]{School of Medical Science and Technology, Indian Institute of Technology Kharagpur, Kharagpur, West Bengal, India - 721302}
\affil[3]{Department of Electrical Engineering, Indian Institute of Technology Bombay, Mumbai, Maharashtra, India-400076}
\affil[4]{Centre for Research in Nanotechnology and Science, Indian Institute of Technology Bombay, Mumbai, Maharashtra, India-400076}
\affil[*]{kasturis@ee.iitb.ac.in}

\begin{abstract}
Wide field-of-view magnetic field microscopy has been realised by probing shifts in optically detected magnetic resonance (ODMR) spectrum of Nitrogen Vacancy (NV) defect centers in diamond. However, these widefield diamond NV magnetometers require few to several minutes of acquisition to get a single magnetic field image, rendering the technique temporally static in it's current form. This limitation prevents application of diamond NV magnetometers to novel imaging of dynamically varying microscale magnetic field processes. Here, we show that the magnetic field imaging frame rate can be significantly enhanced by performing lock-in detection of NV photo-luminescence (PL), simultaneously over multiple pixels of a lock-in camera. A detailed protocol for synchronization of frequency modulated PL of NV centers with fast camera frame demodulation, at few kilohertz frequencies, has been experimentally demonstrated. This experimental technique allows magnetic field imaging of sub-second varying microscale currents in planar microcoils with imaging frame rates in the range of 50 to 200 frames per second (fps). Our work demonstrates that widefield per-pixel lock-in detection of frequency modulated NV ODMR enables dynamic magnetic field microscopy.
\end{abstract}
\begin{document}
\maketitle
\section{Introduction}

The past decade has seen a revolution in high-resolution diffraction-limited microscale and wide field-of-view magnetometry based on optically detected magnetic resonance (ODMR) imaging of Nitrogen Vacancy (NV) defect centers in diamond\, \cite{steiner2010array,pham2011magnetic,lesage,glenn2017micrometer,LevineTurnerKehayiasHartLangellierTrubkoGlennFuWalsworth+2019+1945+1973,tetienne2017quantum}. These room-temperature ultra-sensitive diamond NV magnetometers\,\cite{wolf2015subpicotesla,Barry14133,rondin2014magnetometry,petriniDiamondRev} have enabled a new class of magnetic field microscopy, for example - probing magnetic particles in living cells\,\cite{lesage,davis2018mapping}, imaging fluid-like current flow in graphene\,\cite{tetienne2017quantum,ku2019imaging}, microscopy of novel quantum materials\, \cite{casola2018probing} and rapidly evolving other applications\,\cite{turner2020magnetic,mizuno2020simultaneous,Lillie2019,Broadway2018}.  
In diamond NV-based widefield magnetic field (WMF) imaging, red photo-luminescence (PL) emitted from a microscale volume of NV centers is collected and imaged on to a conventional scientific CMOS or CCD camera. Microwave (MW) resonant frequencies applied to NV centers create changes in NV fluorescence and the precise estimation or tracking of these resonant MW frequencies yields a 2D microscale magnetic field map. The changes in magnetic field experienced by small microscopic volumes of NVs in the diamond crystal get mapped to corresponding pixels on the camera pixel array. 
However, magnetic field images acquired by this method have remained temporally static in nature, demanding few to several minutes of acquisition time for each image frame\, \cite{glenn2017micrometer,davis2018mapping,tetienne2017quantum}. Inherently low NV ensemble resonance contrast and division of informative NV light onto thousands to millions of pixels significantly decrease per-pixel signal-to-noise ratio (SNR) and consequently the magnetic field sensitivity. NV  imaging frame rate for DC to low-frequency magnetometry is fundamentally limited by the NV's optical re-polarization rate i.e. $\sim$ \SI{1}{\mega\hertz}. However, practical SNR bounds have limited imaging frame rates to primarily static magnetic field maps. Development of high-spatially-resolved and high-frame-rate imaging capabilities will enable new applications of NV centers to investigate processes like vortex dynamics in superconductors\,\cite{Lillie2020}, estimating fluctuating magnetic fields from quantum materials\,\cite{casola2018probing}, magnetic nano-particle motion in living cells\, \cite{davis2018mapping,Mahmoudi2011} and imaging mammalian action potential associated magnetic fields\, \cite{barry2016optical,Priceetal20,Webb2021,Parashar2020}.

Detection of weak signals embedded in noise hinges on smart techniques such as the lock-in amplification method, wherein a near-DC or slowly varying signal, mainly submerged in $1/f$ noise, can be periodically modulated and filtered from a narrow band while the noise spanning a large bandwidth can be eliminated leading to significant improvement in signal-to-noise ratio. Pico-Newton scale resolution in atomic force microscopy \cite{SCHLIERF20073989} and high sensitivity magnetometry in SQUIDs and atomic magnetometers \cite{Shah2007} are testament to this detection methodology. With the advent of lock-in cameras\,\cite{1542955}, parallel per-pixel lock-in detection of optical light can be performed over many pixels. In contrast to conventional cameras, the lock-in cameras require synchronized external triggers to perform light integration over specific time windows for each pixel. Intensity measured during these externally timed windows can be used to subtract DC components and estimate the frequency content of the optical signal. With these high frame rate lock-in cameras, new improvements have been observed in techniques where light can be frequency or phase modulated, \textit{e.g.}, deep tissue optical coherence tomography (OCT)\,\cite{lockin_tomography} and ultrasound-modulated OCT\,\cite{liu2016lock} and other avenues\,\cite{Meier2012,SinclairHelicamApp}.  NV's emitted light can be frequency modulated by microwave control of NV resonance \cite{schoenfeld2011real}. Frequency modulated optically detected magnetic resonance (fm-ODMR) schemes for NVs have been used for real-time single point (SP) bulk magnetometry\,\cite{schloss2018simultaneous,clevenson2018robust,barry2016optical,webb2019nanotesla,Webb2021}, where total emitted NV light is collected onto a single photodetector and also for boosting DC-magnetic field sensitivity. A prior work on camera review\,\cite{wojciechowski2018contributed} has also suggested potential application of high-frame rate lock-in camera to perform real-time NV  imaging. 

In this work, we demonstrate a novel per-pixel lock-in detection protocol that enables dynamic millisecond scale magnetic field imaging in wide-field using NV centers in diamond. The paper describes a procedure for synchronizing camera frames of a commercial lock-in camera (Heliotis Helicam C3\,\cite{heliotis_camera}) with NV microwave modulation to obtain fm-ODMR across thousands of pixels. Post calibration of noise statistics and magnetic field sensitivity across different pixels, we measured a median \SI{731}{\nano \tesla / \sqrt{\hertz}} sensitivity per pixel.
To demonstrate spatially and temporally resolved magnetometry, we perform imaging of microscale magnetic fields produced by current flow in two different samples fabricated using e-beam lithography: first, a \SI{10}{\micro\meter} track width gold (Au) microwire with a \SI{90}{\degree} bend and second, a square-spiral planar microcoil of \SI{10}{\micro\meter} track width and full dimensions of \SI{100}{\micro\meter} $\times$ \SI{125}{\micro\meter}. We show dynamic widefield magnetic field images obtained by probing periodically varying current flow in the above samples at near 1Hz , 20Hz and 50Hz magnetic field variations. Multi-pixel fluorescence time traces, scaled to magnetic field values by NV resonance parameters, show expected magnetic field tracking. These sub-second temporal magnetic field images are enabled by fast NV imaging frame rates of 50 to 200 frames per second (fps). To further demonstrate a general application of temporally varying magnetic fields, we show millisecond-scale magnetic field images of current flow in the microcoil from an arbitrary current waveform of varying amplitude and rapid inversion of current direction where the entire event duration is $\approx$\SI{150}{\milli\second}. We discuss the coupling of imaging frame rates and per-pixel SNR to the NV's modulation frequency and the number of signal averaging cycles. Our experimental results demonstrate that frequency-locked widefield imaging of NV emitted light enables dynamic widefield magnetic field imaging at frame rates ranging from \numrange{50}{200}{fps}. Recent work towards dynamic NV widefield imaging \cite{kazi2021wide,hartwalsworth2021}, employ more advanced microwave pulse sequences based on double quantum protocols to significantly reduce heterogeneity in resonant frequencies across imaging field of view which enables high sensitivity magnetic field imaging. In contrast, our results demonstrate high imaging frame rates with a relatively simpler protocol with the application of single resonant MW frequency and could be potentially relevant for a wide variety of NV based imaging applications and can be further improved with spatial homogeneity of resonant frequencies across the field of view.  The scope of the work demonstrated in this paper is not limited to just imaging single crystalline diamonds, but can also be extended to perform improved temporal imaging of nanodiamonds in cellular environments\,\cite{Kucsko2013,Fujiwara,FujiwaraScience2020}.

\section{Experimental Methods}
\subsection{Magnetic Resonance in Nitrogen Vacancy Defects in Diamond }
Negatively charged Nitrogen Vacancy defect centers are point localized Nitrogen substitution of Carbon atoms in the diamond lattice with an adjacent vacancy and an
overall negative charge. Due to the unique electronic properties of these vacancies\, \cite{rondin2014magnetometry}, they are sensitive to external environment changes like, magnetic field, electric field, strain and temperature. The ground state is a spin-triplet with  $m_{s}=0$ and a doubly degenerate $m_{s}=+1$ and $m_{s}=-1$ in the
absence of magnetic field with a zero field splitting of \SI{2.87}{\giga\hertz}. The degeneracy of $m_{s}=+1$ and $m_{s} = -1$ is lifted by
Zeeman splitting in the presence of an external magnetic field. Transitions to the excited state
are spin conserved, however, the relaxation from excited
triplet state take two paths - a radiative spin conserving path and a non-radiative
decay via intersystem crossing (ISCs). The radiative decay produces broadband
red photo-luminescence with the zero-phonon line centered at \SI{637}{\nano \meter}. The non-radiative ISCs are highly
spin-selective towards the $m_s=0$ spin sublevel. Therefore, continuous optical
excitation leads to electron spin polarization. 
Neglecting the hyperfine interaction between the nuclear spin of the nitrogen atom the NV's electronic spin, the ground state NV Hamiltonian is given by 
\begin{equation}
H=h D S_{z}^{2}+h E\left(S_{x}^{2}-S_{y}^{2}\right)+g \mu_{B} B \cdot S,
\label{eq:nv-ground}
\end{equation}
where, $h$ is the Planck's constant, $D$ is the zero-field splitting, $\mu_B$ is the Bohr magneton, $g$ is the gyromagnetic ratio, $E$ is the applied electric field and the last term corresponds to the Zeeman term, with $B$, the externally applied magnetic field. $S_x, S_y, S_z$ correspond to the Pauli matrices for a spin-1 system.
In the weak-field regime where $\mathrm{B}_{\perp} \ll \mathrm{B}_{\|}$, the electron spin resonance frequencies are given by
\begin{equation}
\nu_{\pm}\left(B_{N V}\right)=D \pm \sqrt{\left(\frac{g \mu_{B}}{h} B_{N V}\right)^{2}+E^{2}}
\label{eq:nv_resonance}
\end{equation}
where, $B_{NV}$ is the component of the applied field parallel to the NV axis.
For cases where applied bias field is high enough to neglect the $E$ term, the electron spin resonance frequencies vary linearly with the applied magnetic field. Such a regime is ideal for
sensitive magnetometry with diamond NV centers.

\begin{figure*}[h]
\centering
\includegraphics[width=\textwidth]{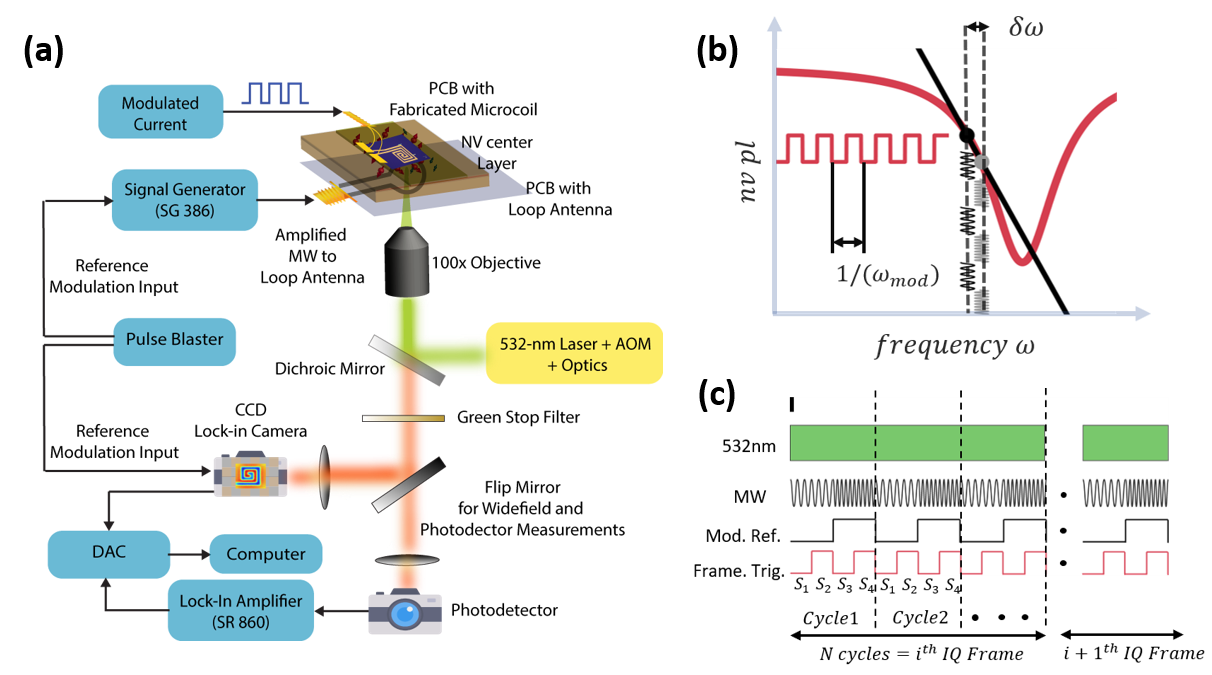}
\caption{Schematic of the experimental setup and protocol for data acquisition: (a) Schematic describing the experimental setup for single photodiode diamond NV magnetometry (SP) or widefield per pixel lock-in diamond nitrogen-vacancy magnetometry (WMF) (b) Illustration explaining generation of frequency modulated NV emitted red light by applying frequency shift key type microwave resonant frequencies. The applied microwave resonant frequencies shuttle between $\omega$ and $\omega-\omega_{dev}$ in sync with square wave waveform of frequency $\omega_{mod}$. When the microwave frequencies are resonant, the emitted NV red light is frequency modulated at $\omega_{mod}$ (c) Pulse protocol to control and synchronize the demodulation of internal camera frames with modulation frequency of optical signal to obtain lock-in in-phase (I) and quadrature (Q) images. A green laser illumination at 532-nm is continuously on and frequency shift key microwave (MW) waveform with modulation $\omega_{mod}$ is applied. Lock-in camera external trigger pulses, controlling internal frame acquisition timings, are provided at $2\omega_{mod}$, synced with MW modulation, where they define 4 quarters for light integration $S_{1} S_{2} S_{3} S_{4}$. These four quarters of light integration allow in-phase ($S_{1}-S_{3}$) and quadrature ($S_{2}-S_{4}$) estimation of optical signal and are averaged over N cycles to give single pair of In-phase image and Quadrature Image (IQ Frame)}  
\label{fig:schem}
\end{figure*}
\subsection{Experimental Setup}
Fig.\,\ref{fig:schem}(a) is an illustration of the experimental setup used to perform diamond NV magnetometry. A non-resonant green light excitation at \SI{532}{\nano \meter} (Sprout Laser)
is used to illuminate NV centers via a $100\times$
objective (Olympus, MLPNFLN series). The excitation beam is focused on the back focal plane of the objective to obtain $\sim$ \SI{200}{\micro\meter} diameter spot size on the NV layer. Optical power impinging the objective back aperture is $\sim$\SI{1.5}{\watt}. We use an isotopically pure diamond crystal (procured from Element Six) of lateral dimensions \SI{4.5}{\mm}$\times$ \SI{4.5}{\mm} and \SI{500}{\micro\meter} thick with a thin \SI{1}{\micro\meter} NV$^-$ implanted layer of 1-2\,ppm \ce{NV-} concentration. The emitted light from NV centers is collected via the same objective,
filtered to select the red light (above \SI{567}{\nano \meter}) and reject green excitation light at (\SI{532}{\nano \meter} using a notch stop filter (SEMROCK NF03-532E-25).
The collected light is focused onto a widefield lock-in camera (Heliotis Helicam C3) to perform widefield magnetometry.
The diamond sample is mounted on a microwave loop PCB,
and associated microwave electronics are used to deliver amplified microwave frequencies
in the range $\sim$ 2.5-3.2\,\SI{}{\giga \hertz}. The applied
microwave frequencies follow frequency shift keying waveforms with square-wave envelopes.
The camera imaging frames are synchronized with the microwave
modulation with specific pulse sequences generated by a high-speed TTL pulse
generator card (SpinCore PulseBlaster ESR-PRO \SI{500}{\mega \hertz}). Samarium-Cobalt (Sm-Co) ring magnets are used for applying a bias magnetic field and have not been shown in the experimental schematic.
Two microscale conductive samples, a $\sim$\SI{10}{\micro\meter} track-width microwire and a $\sim$\SI{10}{\micro\meter} track width planar spiral microcoil was fabricated on two independent silicon substrates. Sample patterning was done using e-beam lithography followed by \SI{100}{\nano\meter} thick deposition of Ti/Au. Typical resistances of these structures were found to be $\sim$\,\SI{400}{\ohm}. These samples are mounted on a custom PCB and wire-bonded to supply the drive voltage. The entire assembly was then glued with Loctite (cyanoacrylate glue) to the diamond crystal. Due to the back focal plane focusing, the 100X objective used in this work suffered a damage in the center of the imaging field of view (FOV) because of high optical intensity localized in a very small area. Consequently, a small number of pixels in the FOV center have zero or minimal ODMR response and can be observed as a small blank hole in all magnetometry related images (for example see Fig.\,\ref{fig:noise}(a) and Fig.\,\ref{fig:static_field}\,(c),(f),(g)).

\subsection{NV Frequency Modulation and Synchronization of Lock-in Camera}
The generation of modulated NV light with frequency $\omega_{mod}$ in the fm-ODMR protocol is shown via the schematic shown in Fig.\,\ref{fig:schem}(b). In a frequency shift keying waveform,
two microwave frequencies $\omega$ and $\omega-\delta\omega$ are delivered via the MW resonator,
where they shuttle between each other with the square wave waveform of frequency $\omega_{mod}$.
For each MW frequency, the NV fluorescence settles to a steady-state value, given by the NV's resonance curve at the applied MW frequency. To measure the amplitude of modulated NV PL, we perform lock-in detection of the collected light at reference frequency $\omega_{mod}$. For the rest of the article, by referring to the 'modulation frequency' of NVs, we also mean the 'reference frequency' of the lock-in camera. 

To synchronize the applied MW waveform with the camera's internal frames, an external reference signal of $2\omega_{mod}$, carefully synced to MW modulation reference at $\omega_{mod}$, is provided to the camera's external trigger input, see Fig.\,\ref{fig:schem}(c). This TTL signal, of twice the modulation frequency, defines the four quarter periods of the
sensor light integration whose values are denoted by $S_{1}, S_{2}, S_{3}, S_{4}$.
The in-phase signal is $S_{1}-S_{3}$ and the quadrature signal $S_{2}-S_{4}$.
Additionally, as shown in the schematic Fig.\,\ref{fig:schem}(c), each cycle of demodulation is internally averaged $N$ times to provide a pair of 2D images containing in-phase (I) and quadrature (Q) values for each pixel. Therefore, to get a single 2D IQ image, the total time is $(1/2\omega_{mod})*cyc$,
which sets the imaging frame rate. Further, since the NV signal scales with $\omega_{mod}$ different imaging frame rates have different SNR as discussed later.
The lock-in camera is limited to frame rates of \SI{3.2}{\kilo \hertz} and a maximum \SI{250}{\kilo \hertz} signal demodulation.     

\section{Results}
\subsection{Optically Detected Magnetic Resonance of Multiple Pixels}

\begin{figure*}[h]
\centering
\includegraphics[width=\textwidth]{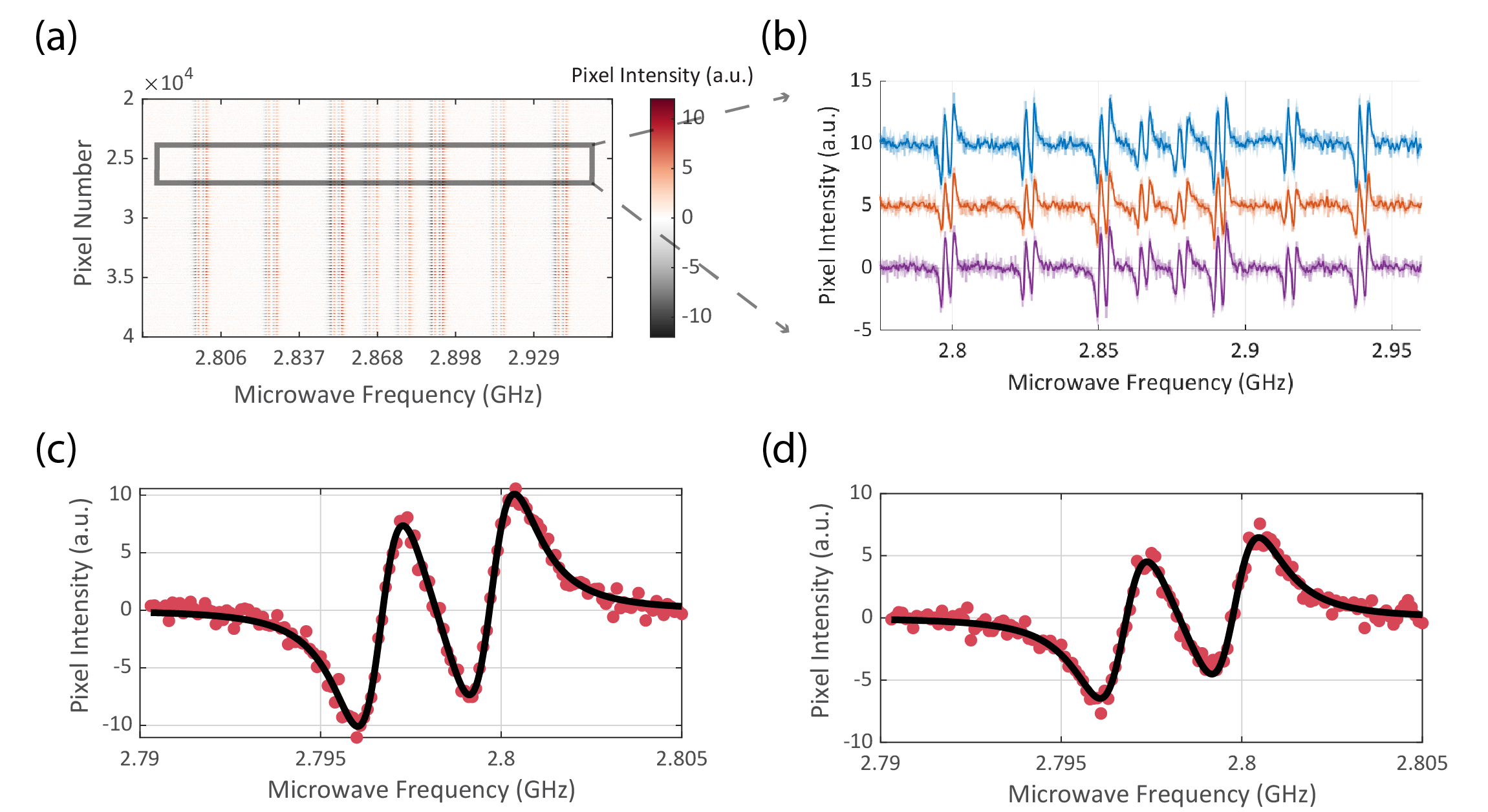}
\caption{Frequency modulated optically detected magnetic resonance spectrum (ODMR) of multiple pixels: (a) A 2D array of $300\times300$ pixels have been concatenated into a 1D array of pixels and their magnetic resonance responses have been color coded. We observe 8 NV resonant frequencies across multiple pixels, with each resonance feature further split into 2 peaks due to N15 hyperfine transitions. (b) Three randomly chosen pixels are used to demonstrate individual pixel ODMR response. The baseline of the pixels, centered at 0, has been shifted to represent them in the same plot. (c) Example pixel ODMR response data recorded at \SI{6.25}{\kilo \hertz} modulation frequency and 122 frame averaging cycles. Each red dot represents data at a single microwave frequency and the black curve represent non-linear Lorentzian-derivative curve fit (d) Example pixel ODMR response data recorded at \SI{8.33}{\kilo \hertz} modulation frequency and 82 frame averaging cycles. Each red dot represents data at a single microwave frequency and the black curve represents non-linear Lorentzian-derivative curve-fit. Reduced ODMR zero-crossing slope can be observed at faster modulation frequencies.}
\label{fig:odmr}
\end{figure*}

Optically detected resonance spectrum of an ensemble of NV centers corresponding to each pixel on the Helicam C3 Array is shown in Fig.\,\ref{fig:odmr}(a). A 2D array of camera pixels have been concatenated into a 1D vector of pixels and their lock-in ODMR response across multiple microwave excitation frequencies have been color-coded. Three randomly selected pixel's individual ODMR traces have been shown in Fig.\,\ref{fig:odmr}(b). For each pixel, the NV response curve can be described by a Lorentzian function,
\begin{equation}
f(\omega)=A\left[1-\frac{C}{ \left[1+\left(\frac{\omega-\omega_{0}}{\Gamma}\right)^2\right]}\right],
\label{eq:lorentzian}
\end{equation}
where $A,C,\Gamma,\omega,\omega_{0}$ denote baseline PL, contrast, the linewidth of resonance,
applied MW frequency, and resonant MW frequency of the NV center respectively.The lock-in signal is proportional to the derivative of the NV ODMR response curve given in Eq.\,\eqref{eq:lorentzian}. 
The derivative of the response curve with an added baseline term was used to fit the lock-in ODMR response, with examples shown in Fig.\,\ref{fig:odmr}(c) and (d). To highlight the importance of NV modulation frequency and frame averaging, two examples of ODMR traces (Fig.\,\ref{fig:odmr}(c) and Fig.\,\ref{fig:odmr}(d)
acquired at different NV modulation frequencies (\SI{6.25}{\kilo \hertz} and \SI{8.33}{\kilo \hertz}) and frame averaging cycles (122 cycles and 82 cycles respectively) are shown. The slope at the zero-crossing point of the fm-ODMR response curve along with the noise floor are critical factors that determine the magnetic field sensitivity of individual pixels. In agreement with previous studies\,\cite{schoenfeld2011real}, we observe reduced zero-crossing slope at higher modulation frequency due to reduced NV interaction time with the
resonant microwave frequencies, oscillating between $\omega$ and $\omega-\omega_{dev}$. The camera pixel readout noise grows with square root of number of the demodulation cycles (HelicamC3 datasheet). This factor introduces a trade-off between the NV response signal and the noise floor with different parameters. Further, the imaging frame rate is dependent on the ratio between modulation frequency to averaging cycles (see Methods, HelicamC3 synchronization), and hence is coupled to the SNR of the NV's ODMR response.

\subsection{Magnetic Field Sensitivity and Static Imaging}
\begin{figure}[h]
\centering
\includegraphics{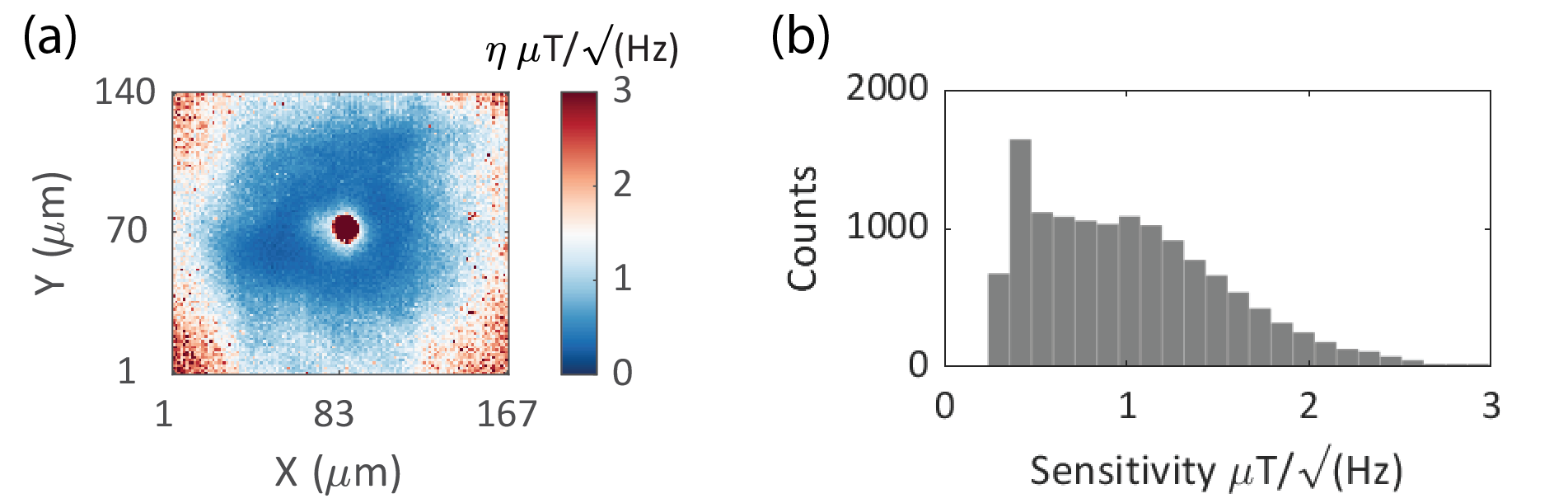}
\caption{Per-pixel sensitivity: (a) Measured 2D map of sensitivity of all responsive pixels. Due to the Gaussian nature of the beam spot, the SNR drops in the outer periphery of the field-of-view(FOV). Pixels with sensitivity better than 3$\mu T/\sqrt{Hz}$ have been included. Some pixels at the center of FOV are non-responsive due to a damage in the objective. (b) Histogram of sensitivity of all responding pixels, with median sensitivity of \SI{731}{\nano\tesla/\sqrt{Hz}}.}
\label{fig:noise}
\end{figure}

As evident from the two example ODMR traces at different acquisition rates, the noise statistics and fm-ODMR signal of pixels can vary significantly with varying image acquisition parameters. Typically, the sensitivity of a sensor is defined by the ratio of uncertainty in the measurement to the maximum slope
point \ie the point of operation of the sensor where the smallest perturbation in the
input creates a maximal change in the output of the sensor. Specifically, for fm-ODMR the slope is maximum at the zero-crossing of the lock-in output, also corresponding to the resonant frequency of NV centers. Therefore, the magnetic field sensitivity is defined as:

\begin{equation}
\eta=\frac{\sigma \sqrt{\tau}}{\left.\frac{d V_{\text {lock }}}{d f}\right|_{V_{\text {lock }}=0, f=\omega_{\text {res }}}}
\label{eq:sensitivity}
\end{equation}
where $\sigma$ is the standard deviation of measurement (voltage for lock-in amplifier or arbitrary units for camera) and
$\tau$ is the measurement time of the signal and \textit{f} is the frequency. The denominator denotes the slope at the resonant frequency $\omega_{res}$. 

To acquire the $\sigma$ for individual pixels, sixty imaging frames were acquired and the mean and standard deviation of each pixel's intensity were recorded. Example noise spectrum of WMF pixels as a function of lock-in modulation frequencies have been shown in Supplementary Fig.\,\ref{fig:hc3_noisespectrum}(b),(c) along with a typical $1/f$ noise spectrum of a single-photodiode (SP) lockin measurement (Supplementary Fig.\,\ref{fig:hc3_noisespectrum}(a). The WMF $\sigma$ spectrum for most pixels remained approximately flat, as compared to the SP $\sigma$ spectrum, between modulation frequencies of 3-100\,\SI{}{\kilo\hertz}, with mean value of 1.95 units (out of 10-bit 1024 point scale) for all pixels in Fig.\,\ref{fig:hc3_noisespectrum}(c). Since the minimum possible camera modulation frequency is \SI{2.2}{\kilo \hertz}, most of the low-frequency noise is eliminated in the WMF noise spectrum. For WMF imaging experiments $\tau=(1/\omega_{mod})*n_{cyc}$, where $n_{cyc}$ is the number of frame averaging cycle. To measure the zero-crossing slope an ODMR spectrum is measured with a frequency resolution of (\SI{100}{\kilo \hertz}). The slope at
zero-crossing for each pixel is then obtained by non-linear curve fitting and the corresponding 2D sensitivity map is shown in Fig.\,\ref{fig:noise}(a), depicting a spatial variation of pixel sensitivity by evaluating Eq.\,\eqref{eq:sensitivity} for each pixel. The pixel response mimics the excitation profile. As expected, pixels with high response, fall within the central region of the FOV and pixels with low or no response, fall towards the outer periphery of the NVs PL intensity profile. The distribution of per-pixel sensitivity is shown in Fig.\ref{fig:noise}(b) where a median pixel sensitivity of \SI{731}{\nano \tesla / \sqrt{\hertz}} is observed. Only pixels with sensitivity more than \SI{3}{\micro\tesla\per\sqrt{\hertz}} have been considered due to the low ODMR response at the outer periphery of the beam. Additionally, before the curve fitting for each pixel, a selection threshold was applied to select pixels with a minimum threshold level of fm-ODMR response (see Supplementary notes, per-pixel raw data processing) and only the responding pixels were further analyzed.

\begin{figure}[h]
\centering
\includegraphics{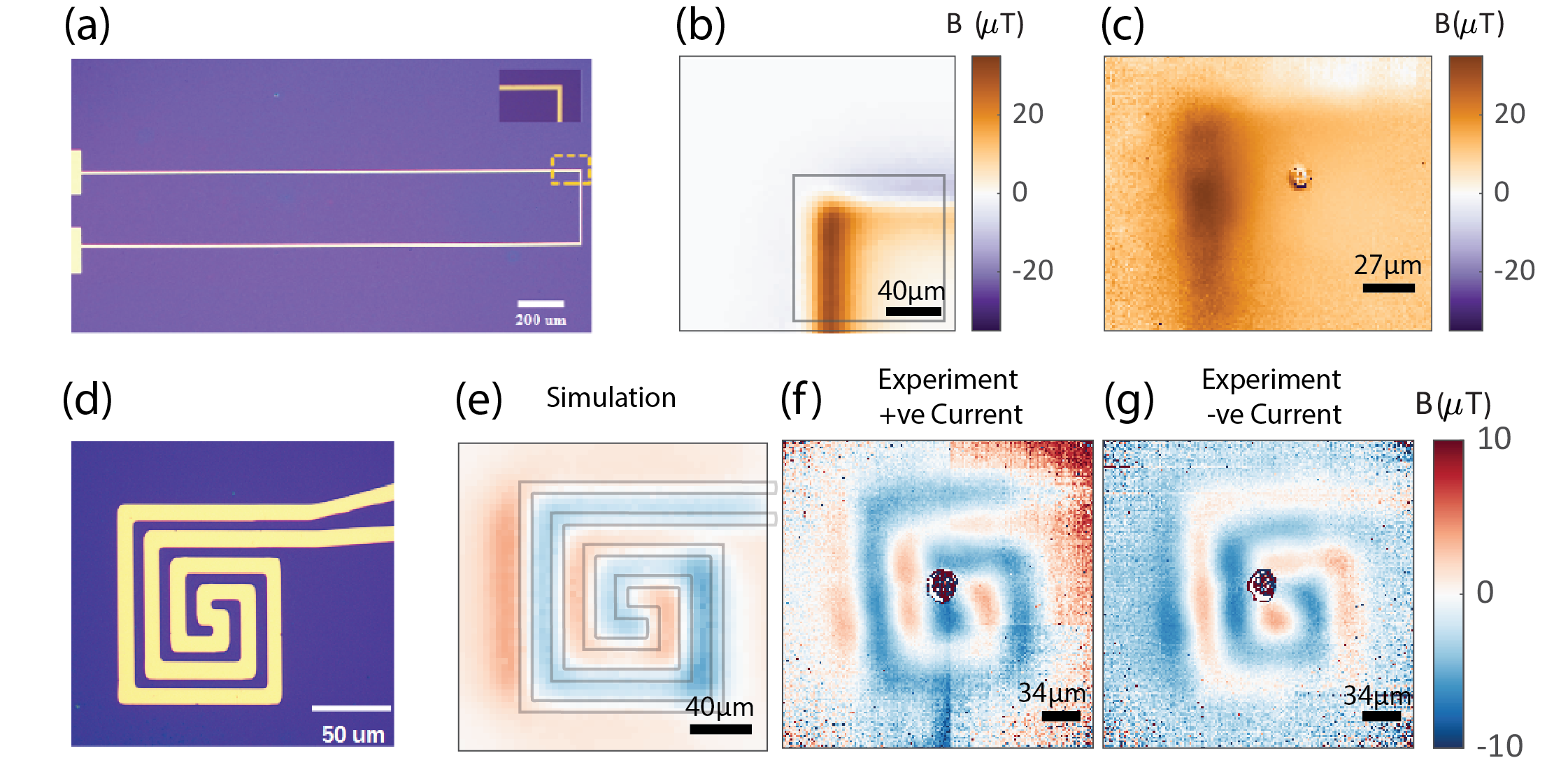}
\caption{Static magnetic field images of the microwire and the microcoil sample: (a) Color microscope image of the U shaped microwire sample. Microwire track width is \SI{10}{\micro\meter}. Scale Bar \SI{200}{\micro\meter}. Inset shows the \SI{90}{\degree} bend feature which has been imaged. (b) Simulation of single NV-axis magnetic field map of the \SI{90}{\degree} bend feature of the microwire, at a standoff \SI{13}{\micro\meter} and current \SI{2.4}{\milli\ampere}. Scale bar \SI{40}{\micro\meter}. Black square indicates the approximate NV magnetic field imaging field of view location. (c) Experimentally measured magnetic field image of the microwire with \SI{2.4}{\milli\ampere} current flow, about the same NV axis as shown in simulation. Scale bar \SI{27}{\micro\meter}. (d) Color microscope image of the microcoil sample with metal track width \SI{10}{\micro\meter} and overall dimensions \SI{100}{\micro \meter}$\times$ \SI{125}{\micro \meter} . Scale bar \SI{50}{\micro\meter}. (e) Simulation of the single NV-axis magnetic field map of the microcoil, at standoff \SI{14}{\micro\meter} and \SI{500}{\micro\ampere} current flow. Sample geometry, translucent gray lines, has been scaled to simulation field image and overlaid for easy comprehension of the current flow path. Scale Bar \SI{40}{\micro\meter}. (f) Experimentally obtained single axis magnetic field image of the microcoil for positive direction \SI{500}{\micro\ampere} current flow, about the same NV axis as shown in simulation. Scale Bar \SI{34}{\micro\meter}. (g) Experimentally obtained single axis magnetic field image of the microcoil for negative direction \SI{500}{\micro\ampere} current flow, about the same NV axis as shown in simulation. Scale Bar \SI{34}{\micro\meter}.} 
\label{fig:static_field}
\end{figure}


Spatial and temporal resolutions are inherently coupled in diamond NV microscopy. We verify magnetic field image formation with static acquisition (5-10 minutes) for two microscale samples, one \SI{10}{\micro\meter} track width microwire and one spiral microcoil of \SI{10}{\micro\meter} track width and overall dimensions of \SI{100}{\micro \meter}$\times$ \SI{125}{\micro \meter} , as described earlier in methods. The two sample images are shown in Fig.\,\ref{fig:static_field}(a) and Fig.\,\ref{fig:static_field}(d). Magnetic field images, projected onto a single NV axis, of these samples were formed by \SI{100}{\kilo\hertz} step size sampling of the NV resonance, non-linear parameter fits for individual pixels and subsequent determination of a map of resonant frequencies for 2D array of pixels. The resonant frequency maps of these samples were acquired for both DC current on and off and subtracted to probe sample magnetic field dependent on linear shifts in the resonant frequencies. Single NV axis magnetic field images of these samples (Fig.\,\ref{fig:static_field}(c) microwire and Fig.\,\ref{fig:static_field}(f),(g) microcoil) were in agreement with simulated magnetic field images obtained using COMSOL Multiphysics, ( Fig.\,\ref{fig:static_field}(b) microwire and Fig.\,\ref{fig:static_field}(e) microcoil for expected simulated field images) at an estimated standoff of $\sim$ \SI{13}{\micro\meter} for the microwire and $\sim$\SI{14}{\micro\meter} for the microcoil.    

The $\textbf{B}_{NV}$ field is measured by resonant frequency shifts on either side of a reference bias field resonant frequency. Therefore, on inverting the current direction in the sample we observed an inverted contrast in the magnetic field image of the microcoil sample as shown in Fig.\,\ref{fig:static_field}(f) for arbitrarily defined positive current and Fig.\,\ref{fig:static_field}(g) for negative current, which further affirms that the magnetic field images obtained are from the microscale current flow in the sample. Additionally, static acquisition allows for quantification of the field of view, the spatial resolution of the imaging setup and the effective magnification. The imaging field of view, with sufficient NV resonance SNR, is $\sim$ \SI{150}{\micro\meter} $\times$ \SI{150}{\micro\meter} (Fig.\,\ref{fig:noise}) and is limited by the excitation beam spot size on the NV layer and the total optical power of the Gaussian excitation \SI{532}{\nm} beam in our experimental setup (with $\sim$\SI{1.5}{\W} entering the objective back aperture). We estimated the spatial resolution to be \SI{1.7}{\micro\meter} per camera pixel (see Supplementary note for pixel resolution estimation method) during microcoil measurements and \SI{1.33}{\micro\meter} per camera pixel during microwire measurements. Spatial resolution slightly differs in the two measurements due to minor differences in positioning of a focusing plano-convex lens in the red emitted light collection path to incorporate a larger field of view for the microcoil. Consequently, corresponding effective magnifications were 30X for microwire measurements and 23.5X for microcoil measurements in our widefield microscope.
While we acquire only single NV axis magnetic field static and dynamic images in this study, we show that the microcoil sample's vector magnetic field can be reliably reconstructed (Supplementary Fig.\,\ref{fig:fourier_reconstruction}) from single NV axis magnetic field images by well established Fourier reconstruction methods \,\cite{Lima2009ObtainingVM,glenn2017micrometer}.  

\subsection{Dynamic Widefield Magnetic Imaging}

In this section we describe the acquisition of millisecond scale widefield magnetic field images. To perform real-time imaging, the applied microwave frequency is fixed to a specific NV resonant frequency along one NV axis. An externally applied magnetic field causes a linear shift in pixel intensity, proportional to the zero-crossing NV slope. Therefore, tracking the pixel intensities, scaled by the slope, gives a measure of the external magnetic field fluctuation along the chosen NV axis corresponding to each pixel. The time-dependent magnetic field can be estimated from

\begin{equation}
B(t)=\frac{v(t)-v_{o}}{\left.\frac{d V_{\text {lock }}}{d f}\right|_{V_{\text {lock }}=0, f=\omega_{\text {res }}}}\gamma
\label{eq:tracking}
\end{equation}

where $v(t)$ is the lock-in pixel intensity, $v_{o}$ is a fixed offset baseline of the pixel and $\gamma = $  \SI{28}{\kilo \hertz \per  \micro \tesla} is the gyromagnetic ratio. The zero-crossing slope scale factor is independently determined corresponding to each pixel in the imaging window. Individual pixels are heterogeneous in their resonant frequencies due to small deviation arising from local crystal strain, non-uniform bias magnetic field and temperature in the excitation volume of the diamond sample. Therefore, we choose to select the median resonant frequency from the distribution of resonant frequencies in the imaging window for widefield magnetic field tracking. 

We demonstrate temporal magnetic field imaging examples for both samples, the microwire and the microcoil, at different magnetic field variations and imaging frame rates. The current flow in these samples are controlled by an arbitrary waveform analog voltage generator (NIDAQ PCIe-6363, Analog output) and the applied voltage waveform is triggered in synchronization with camera frame acquisition (see Fig.\,\ref{fig:schem}). A low peak current level of \SI{500}{\micro\ampere} was chosen for temporal field imaging demonstration of both samples to keep peak magnetic field values below $\sim$ \SI{6}{\micro\tesla} in the entire FOV, at the given sample-standoff (see static imaging section and Fig.\,\ref{fig:static_field}).

\begin{figure}[t]
\centering
\includegraphics{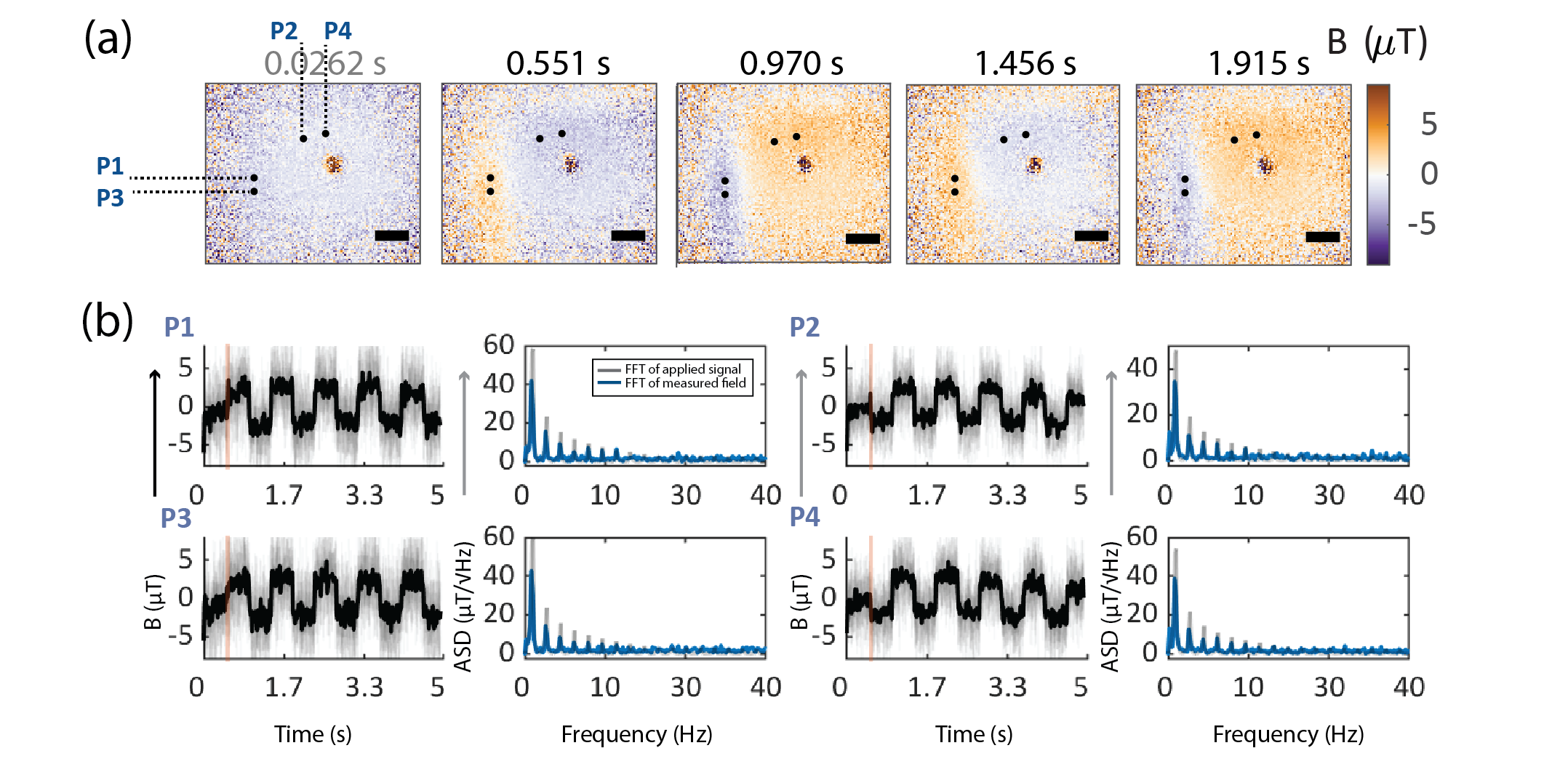}
\caption{Temporal imaging of 1.26 Hz magnetic field variation at 78 frames per second of the \SI{90}{\degree} bend feature of the microwire sample: (a) Magnetic field frames at single time-points (averaged $n=15$ iterations) showing alternating field image contrast with reversal in current direction. No voltage applied for a baseline time of \SI{0.5}{\second}, first frame selected from baseline window. A periodic square wave voltage waveform of alternating polarity was applied after the baseline time, at \SI{1.26}{\hertz} periodicity and peak current \SI{500}{\micro\ampere}. Exact magnetic field frame time-points have been shown on top of each image. Scale bar \SI{27}{\micro\meter} (b) Full temporal time traces of example pixels showing magnetic field tracking in time. Location of example pixels, labelled as P1 to P4 in the field of view have been shown in the magnetic field images. For each pixel, magnetic field traces versus time show tracking of applied the magnetic field, with faded gray lines as single iteration traces and solid black lines showing mean ($n=15$) magnetic field traces for the given pixel. Amplitude spectral density of single-pixel field traces are shown on the left, where pixel Fourier spectra are in blue and applied voltage Fourier spectra has been shown in gray. Applied voltage spectral density is scaled to a constant to compare spectral content with pixel Fourier spectra. Since pixels track magnetic field, peaks in the pixel Fourier spectra matches with peaks in the Fourier spectrum of the applied voltage, with peaks occurring at magnetic field variation \SI{1.26}{Hz} and it's odd harmonics.} 
\label{fig:temporal1}
\end{figure}

\textbf{Microwire imaging, \SI{1.26}{\hertz} sample field variation, 78 fps NV acquisition:} Dynamic magnetic field imaging was performed on the microwire sample, where acquisition rate of magnetic field frames was set to 78 fps and a \SI{1.26}{\hertz} periodic square bipolar voltage waveform was applied to the microwire. A peak current of \SI{500}{\micro\ampere} produced peak magnetic field around \SI{5}{\micro\tesla} in the imaging FOV. Fig.\,\ref{fig:temporal1}(a) shows example single magnetic field frames (iterations $n=15$) at selected time-points demonstrating temporally varying, alternating field magnetic image contrast due to periodic changes in current polarity. Few example pixels P1 to P4 (see Fig.\,\ref{fig:temporal1}(b)) have been selected to show full temporal response of these individual pixels. Single-iteration time traces (Fig.\,\ref{fig:temporal1}(b), faded gray traces from all $n=15$ iteration) and  mean time traces ($n=15$, Fig.\,\ref{fig:temporal1}(b), black solid traces) of individual pixels track applied magnetic field waveform. Fourier spectra of these pixel time traces are observed to contain peaks at odd harmonics of applied magnetic field variation \SI{1.26}{Hz} as expected. Example pixels P2 and P4 were selected at perpendicular locations to the current path near pixels P1 and P3. Therefore, we observe P2 and P4 time traces are $\sim$ \SI{90}{\degree} phase shifted to P1 and P3 time traces, inline with expected spatial magnetic field profile of the microwire. To the best of our knowledge this demonstrates the first real-time tracking of microscale magnetic fields which faithfully reconstructs the frequency and phase of the applied field. A link to the video file of this imaging dataset has been provided in the supplementary section.

\begin{figure}[h]
\centering
\includegraphics[width=0.95\textwidth]{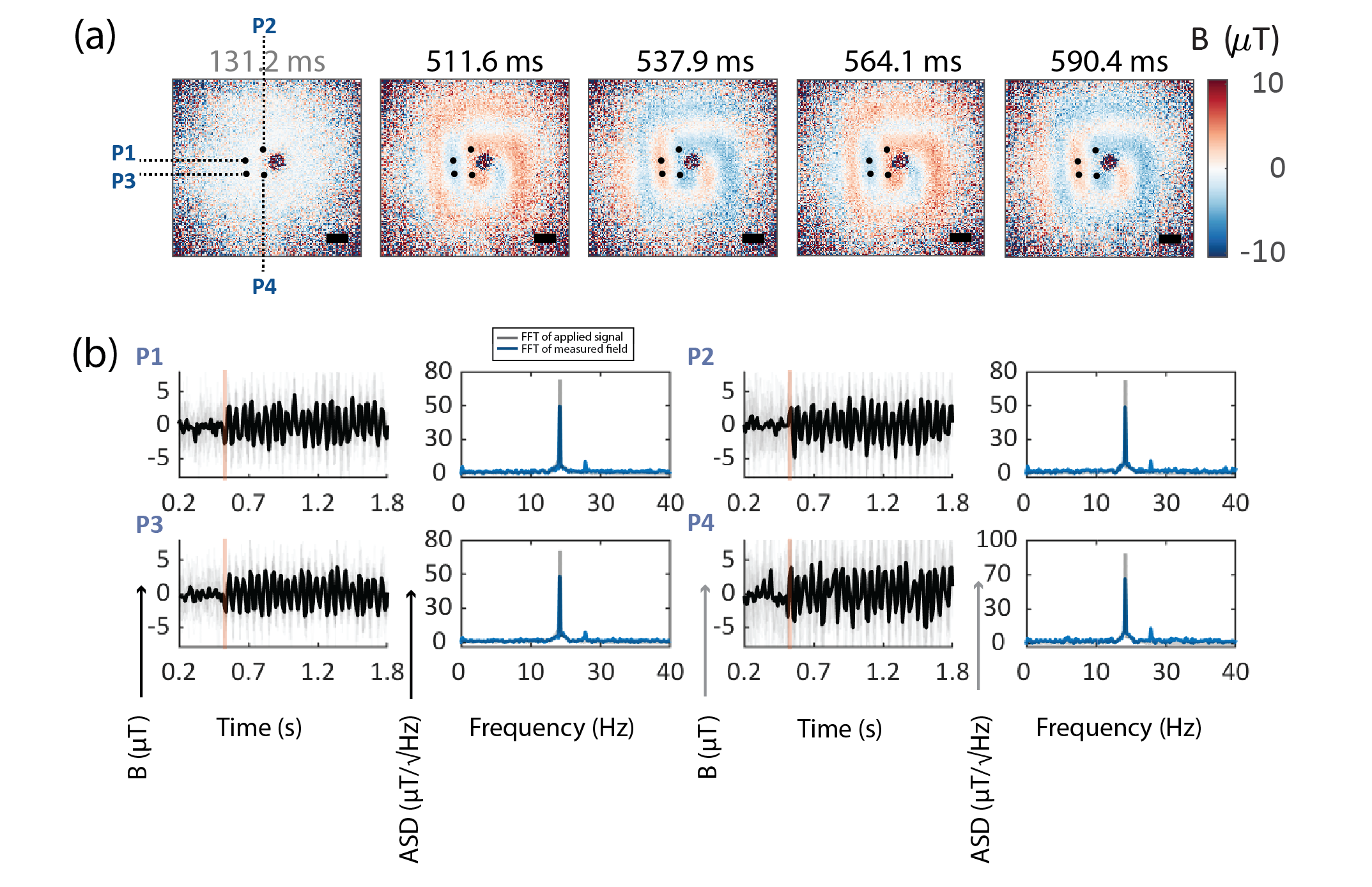}
\caption{Temporal imaging of \SI{17.9}{\hertz} magnetic field variation at 78 frames per second of the microcoil sample: (a) Magnetic field frames at single time points (averaged $n=15$ iterations) showing alternating field image contrast with reversal in current direction. No voltage applied for a baseline time of \SI{0.5}{\second}, first frame selected from baseline window. A periodic square wave voltage waveform of alternating polarity was applied after the baseline time, at \SI{17.9}{\hertz} periodicity and peak current \SI{500}{\micro\ampere}. Exact magnetic field frame time points have been shown on top of each image. Scale bar \SI{34}{\micro\meter} (b) Full temporal time traces of example pixels showing magnetic field tracking in time. Location of example pixels, labelled P1-P4 in the field of view have been shown in the magnetic field images. For each pixel, magnetic field traces versus time show tracking of applied the magnetic field, with faded gray lines as single iteration traces and solid black lines showing mean ($n=15$) magnetic field traces for the pixel. Amplitude spectral density of single-pixel field traces are shown on the left, where pixel Fourier spectra are in blue and applied voltage Fourier spectra has been shown in gray. Applied voltage spectral density is scaled to a constant to compare spectral content with pixel Fourier spectra. Since pixels track magnetic field, the prominent peak in the pixel Fourier spectra matches with the peak in the Fourier spectrum of the applied voltage, both occurring at magnetic field variation \SI{17.9}{\hertz}.} 
\label{fig:temporal2}
\end{figure}

\textbf{Microcoil imaging, \SI{17.98}{\hertz} sample field variation, 78 fps NV acquisition:} Dynamic magnetic field imaging was performed on the planar microcoil sample, where the NV acquisition rate was set to 78 fps and a \SI{17.98}{\hertz} periodic square voltage waveform  was applied to the microcoil. Results of microcoil imaging (see Fig.\,\ref{fig:temporal2}) have been similarly organized as discussed in the microwire temporal imaging text. Microscale magnetic field profiles of the microcoil are spatially resolved in single sub-second magnetic field frames (Fig.\,\ref{fig:temporal2}(a), 12ms per frame, $n=15$). Magnetic field time traces of example pixels have been shown in Fig.\,\ref{fig:temporal2}(b) and example pixel locations on the microcoil images have been marked in Fig.\,\ref{fig:temporal2}(a). Fourier spectra of these pixels show peak at the frequency of applied \SI{17.98}{Hz} periodic magnetic field waveform. These results demonstrate resolving spatially intricate field profiles, in this case multiple current flow paths separated by $\sim$\SI{7}{\micro\meter}, at millisecond scale snapshots of magnetic field images. A link to the video file of this imaging dataset has been provided in the supplementary section.

Temporal imaging data for the microcoil at similar magnetic field variation \SI{18.9}{\hertz}  but higher NV acquisition rate of 208 fps has been shown in the supplementary section (Supplementary Fig.\,\ref{fig:supple_uspiral_20Hz208fps}). Microcoil magnetic field features are spatially-resolved with reduced SNR and the first odd harmonic of \SI{18.9}{\hertz} field variation is also observed in the Fourier spectra of individual pixel responses. A higher magnetic field variation \SI{41.52}{\hertz} applied to the microcoil at 208fps acquisition rate is also shown (Supplementary Fig.\,\ref{fig:supple_uspiral_41Hz208fps}). Additionally, for completeness, we show dynamics in the microwire sample at similar magnetic field variations (\SI{16.3}{\hertz}) and 78fps NV acquisition rate. Supplementary Fig.\,\ref{fig:supple_uwire_29Hz78fps} shows spatially resolved magnetic field images of the microwire and expected magnetic field tracking in individual pixel responses.

\begin{figure}[t]
\centering
\includegraphics[width=\textwidth]{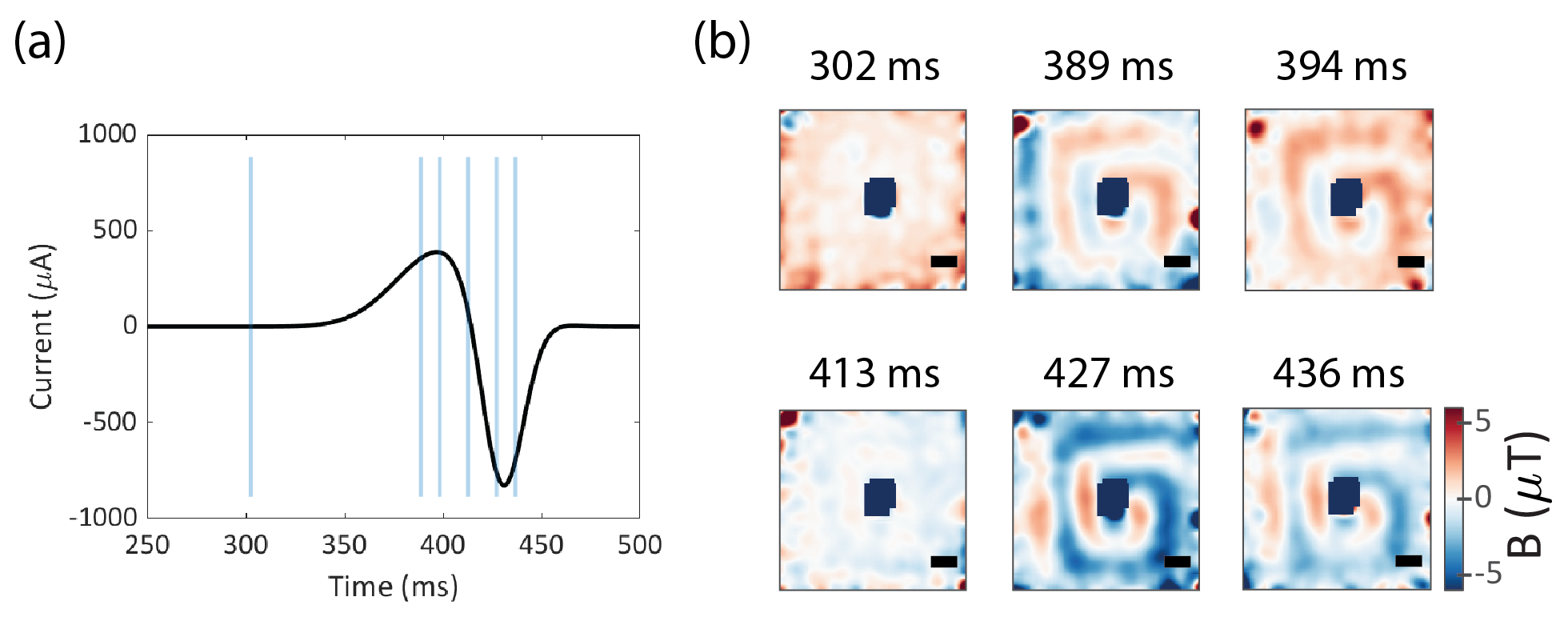}
\caption{Temporal imaging of an arbitrary millisecond scale magnetic field variation at 208 frames per second of the microcoil sample: (a) Applied current profile to the microcoil sample. The main waveform signature lasts for less than \SI{150}{\milli\second}. Vertical blue lines indicates time-points where single magnetic field image frames have been shown further. (b) Example magnetic field frames at selected time-points (averaged $n=15$ iterations) have been shown. Magnetic field images are Gaussian-smoothened with 4.5$\sigma$ filter. The applied current profile is reflected in the series of spatially resolved magnetic field images of the microcoil. Magnetic field image at \SI{413}{\milli\second} is observed to faithfully capture the fast inversion of current polarity. Scale bar \SI{34}{\micro\meter}.} 
\label{fig:APliketemporal}
\end{figure}

\textbf{Arbitrary waveform dynamics:}
Here, we show millisecond scale widefield magnetometry for a generalized arbitrary waveform signal (Fig.\,\ref{fig:APliketemporal}a) with a rapid inversion of current polarity and where the total fluctuation event lasts less than 150ms. Fourier spectrum of this applied waveform contains energy in frequencies upto \SI{100}{\hertz}. Therefore, to sufficiently sample the magnetic field profile, number of demodulation cycles were reduced to get NV acquisition rate of 208fps. Fig.\,\ref{fig:APliketemporal}(b) shows selected single magnetic field frames (\SI{4.8}{\milli \second} NV acquisition time per frame, $n=15$ iterations) that show expected microcoil field temporal profile in response to the applied waveform. Notably, field frame at \SI{413}{\milli\second} captures the point of near-zero magnetic field profile when the current rapidly switches polarity within $\sim$\SI{30}{\milli\second}. Since magnitude of the peak negative current is higher than the peak positive current, microcoil features are more prominent in field frame at \SI{427}{\milli\second} as compared to field frame at \SI{394}{\milli\second}. Magnetic field images for this case have been Guassian-smoothened with $4.5\sigma$ filter to remove additional noise in temporal images incorporated at higher imaging frame rates. A link to the video file of this imaging dataset has been provided in the supplementary section.

Further, we observed high frequency noise in lock-in camera pixel response which can be reduced by the use of appropriate filtering techniques like Bayesian filtering to further enhance imaging SNR. To the best of our knowledge, the acquired single-axis widefield magnetic field images constitute a novel demonstration of real-time millisecond scale widefield magnetic field microscopy. Improved temporal resolution is primarily enabled by pixel noise-rejection at higher lock-in frequencies, high imaging frame rates offered by the lock-in camera and ability to synchronize modulation of NV emitted light with lock-in camera frame 
integration timings. At high imaging frame rates, the SNR is primarily limited by the NV's emitted 
fluorescence rate from the diamond sample, and not by the lock-in camera demodulation rates. Therefore, the temporal imaging enhancement demonstrated in this work is expected to improve at least one-two fold with optimized optical and microwave excitation power of the NV ensemble and further, by the use 
state-of-art ion-irradiated high density nitrogen vacancy diamond samples.

\textit{Imaging speed and sensitivity trade-off}: Finally, we discuss the interplay of four key parameters of WMF imaging method, namely, the imaging frame rate $I$, the mean per-pixel sensitivity $\eta$, the NV modulation frequency $\omega_{mod}$ and the number of frame averaging cycles $n_{cyc}$. A phenomenological understanding of the coupling of parameters will be useful in deciding the trade-off. To maximize the imaging frame rate $I$ $\propto \omega_{mod}/n_{cyc}$, we need to modulate NVs faster (increase $\omega_{mod}$) and average for lesser number of internal frames (decrease $n_{cyc}$). Increasing $\omega_{mod}$ leads to a decrease in the zero-crossing NV slope
but the noise, $\sigma$, remains mostly constant. Therefore, $\eta$ will drop at higher 
$\omega_{mod}$, keeping $n_{cyc}$ same. Increasing the $n_cyc$ has more
interesting effects on $\eta$, since the camera readout noise $\sigma$ increases
with more $n_{cyc}$ but the NV signal strength also improves. Therefore, a multi-parameter optimization is required for understanding the trade-offs and zone of best performance for the sensor for a given specific application.

In summary, we have developed a novel widefield magnetic field microscope capable of probing dynamically varying microscale magnetic field features at tunable imaging frame rates of 50-200 frames per second. Millisecond to sub-second magnetic field images have been demonstrated for a planar microcoil sample with detailed microscale features, consisting of multiple current flow paths separated by $\sim$ \SI{7}{\micro\meter}, current flow track width \SI{10}{\micro\meter} and multiple \SI{90}{\degree} turns in the current flow path. While maintaining microscale spatial resolution, individual pixels in the imaging FOV have been shown to track applied magnetic fields in time with correct amplitude and phase for both periodic current waveforms and short \SI{150}{\milli\second} arbitrary current waveform. Frequency spectrum of individual pixels reveal near exact match to the frequency spectrum of applied periodic current waveforms. Further, the NV imaging speed enhancements have been shown for small magnetic fields, typically less than $\sim$\SI{6}{\micro\tesla} in the entire FOV. Therefore, to the best of our knowledge, the widefield per-pixel lock-in method proposed here marks significant improvement over conventional ODMR imaging, where few to several minutes of averaging time is required to obtain a single magnetic field image of similar microscale spatial resolution.  


\section{Conclusion and Outlook}

In this work, we have developed and demonstrated an experimental technique to perform real-time widefield magnetic field imaging using diamond NV centers in diamond. Per-pixel SNR is significantly enhanced using lock-in detection techniques implemented on a commercial lock-in camera which allows simultaneous demodulation of multiple pixels. While previous diamond NV based magnetometers have shown acquisition rate of several minutes per frame, to the best of our knowledge, we demonstrate for the first time, spatio-temporal magnetic field imaging at timescale around 1-40\,\SI{}{\hertz} at imaging speed of \numrange{50}{200}\,fps. The fm-ODMR protocol used in this demonstration is easy to implement, demanding only frequency modulated NV-PL and microsecond digital pulses that control camera frame demodulation. We expect temporal imaging SNR and imaging FOV shown in our work to significantly improve with increase in optical excitation of NV centers and with application of state-of-art higher NV density diamond crystals. Additionally, the spatio-temporal resolution is expected to improve in future with the use of higher NV concentration diamond samples and improved coherence time. We emphasize that while we operate the camera at demodulation of 6.25-8.33\,\SI{}{\kilo \hertz} and imaging frame rates of $\sim\, 50-200$ fps, the demonstration is primarily limited by the low NV fluorescence and not by maximum achievable lock-in modulation rates (possible up to \SI{250}{\kilo \hertz}) and imaging frame rates (maximum possible 3200 fps) for the camera used here. Other lock-in cameras\,\cite{Cao:20,lockin_phase} are expected to offer similar high frame rate advantages. We are aware of a similar independent preprint submission by Webb\,\etal\,\cite{webb2021high} where the authors demonstrate an application of widefield lock-in detection to enhance imaging speed of diamond NV magnetometry. Both our work and their work, with differences in experimental implementation, show that widefield lock-in detection enables sub-second magnetic field microscopy using NV defect centers in diamond, in contrast to conventional static diamond NV magnetic field microscopy.

\section*{Acknowledgements}

K.S. acknowledges financial support from IIT Bombay seed grant number 17IRCCSG009, DST Inspire Faculty Fellowship - DST/ INSPIRE/04/2016/002284, SERB EMR grant Number EMR/2016/007420 and Asian Office of Aerospace Research and Development (AOARD) R$\&$D grant No. FA2386-19-1-4042. K.S. acknowledges the support and usage of fabrication facilities in the IIT Bombay Nanofabrication facility via the NNetra project sponsored by Department of Science and Technology (DST) and Ministry of Electronics and Information Technology (MEITY), India. This work was also supported by the DBT/Wellcome Trust India Alliance Fellowship IA/I/11/2500270 awarded to S.B.. M.P. thanks MHRD, India for Institute Fellowship and Prime Minister's Research Fellowship (PMRF). The authors thank Heliotis Helicam technical assistance, especially Istvan Biro, for his help in camera synchronization. K.S. acknowledges the contribution of Aditya Malusare and Parth Jatakia during the initial experimental setup.  The authors thank Dr. Siddharth Tallur for allowing the usage of SR860 Lockin amplifier for single-photodiode experiments and also thank Prof. Pradeep Sarin, Prof. Kantimay Dasgupta and Bikas C.Barik for access and help in wire bonding the micro-fabricated samples. The authors note that the work has been provisionally filed under the Indian Patent Act with application number:202121010532.

\section*{Author contributions statement}

M.P, S.B, and K.S conceived the idea. M.P and K.S designed the experimental setup. M.P constructed the experimental setup, wrote custom software for experiment control and data acquisition and performed all primary experiments and data analysis. A.B designed and performed micro-fabrication of the microwire and the microcoil sample. A.B simulated magnetic field profiles of the samples. D.S., A.B and A.G assisted data collection, designed and characterized microwave loop PCB. D.S, A.B and S.B contributed key ideas to experiments and data analysis. M.P and K.S. wrote the manuscript in discussion with S.B . All authors reviewed and approved of the manuscript. K.S supervised all aspects of the work.

\section*{Competing Interests}
The authors declare no competing interests.

\renewcommand\thesection{S\arabic{section}}
\setcounter{section}{0}
\renewcommand\thefigure{S\arabic{figure}}    
\setcounter{figure}{0}    

\section{Supplementary information}

\subsection{Supplementary notes}

\textbf{Normalization of amplitude spectral density:} Amplitude spectral density (ASD) of magnetic field traces of individual pixels is defined as the square root of one-sided power spectral density (PSD). One-sided PSD $S(f)$ of a pixel time series X(t) is normalized such that area under the PSD curve integrated on one-side from 0 to $\frac{f_{s}}{2}$ equals the variance $\sigma_X^2$ of the mean subtracted pixel time traces. $f_{s}$ denotes sampling frequency or frames per second.   

\begin{equation}
\sigma_X^2 = \int_{0}^{\frac{f_s}{2}} S(f)\dd{f} 
\label{eq:asd}
\end{equation}

The above normalization was implemented with MATLAB in-built functions. For a time series data vector $V$, the ASD vector is given by 
\begin{verbatim}
fourier_vector = fftshift(fft(V))
ASD = sqrt(2) * abs(fourier_vector) / sqrt(length(fourier_vector))
\end{verbatim}

The right half of the ASD vector represents amplitude spectral density from frequency 0 to $f_s/2$.

\textbf{Determination of single pixel spatial resolution and effective magnification:} In the experimental setup, the entire assembly of sample, diamond crystal and microwave resonator was mounted on a motorized XYZ stage and the excitation beam was kept fixed. The motorized stage coordinates are accurate to \SI{100}{\nano\meter} positioning. Magnetic field images of the microcoil sample were acquired at slightly shifted ($\sim$ \SI{20}{\micro\meter}) locations in X and Y motor coordinates. Corresponding to the change in motor coordinates, number of pixel shifts were noted for a sharp feature in the magnetic field image of the microcoil or the microwire sample. The measurements were repeated several times and the per-pixel spatial resolution was evaluated to be \SI{1.33}{\micro\meter} per pixel during the microwire measurements and \SI{1.7}{\micro\meter} per pixel during the microcoil measurements. As mentioned in the main text, the per-pixel resolution during the two sets of sample measurements differ due to slight change in positioning of a focusing plano-convex lens in the excitation-fluorescence collection path of the widefield microscope. The lock-in camera real pixel size is \SI{40}{\micro\meter}, which yields effective magnification of $30\times$ for microwire measurements and $23.5\times$ for microcoil measurements.       

\textbf{Per-pixel raw data processing:} 
Additional details to measure raw-data, process and analyze time-dependent magnetic field maps.

\begin{enumerate}
\item Before dynamic magnetic field tracking, we acquire a widefield lock-in ODMR spectrum of a single NV resonant feature at high microwave frequency step size of \SI{100}{\kilo \hertz}. The resonant feature is selected on the basis of high signal response and linearity at the NV zero-crossing point \ie the NV resonant frequency. This selection determines the NV axis along which the magnetic field sensing will be performed.
\item The informative red light emitted from NV centers spans a limited area on the CMOS array ($300\times 300$) pixels. Further, the NV ODMR signal of different pixels differ due to Gaussian nature of optical illumination, spatial non-uniformity of the applied microwave field and limited spot size of the excitation beam. Therefore, it is important to select responding pixels. First we create an average response template by taking mean of ODMR response of all pixels, responding and non-responding pixels. Since a high number of pixels are responsive in the ODMR data, the template carries an average ODMR feature. The template is normalized to unit norm and the unit-norm responses of all pixels are correlated to the template, via the dot product. Pixels with projection values higher than a set threshold are selected for further processing.  This threshold was kept low at 1e-4 to only reject extremely low SNR pixels. Additional selection of high response pixels occurs with subsequent process of non-linear curve fitting.   
\item Non-linear curve fitting is performed to fit derivative sum of two Lorentzian profiles separated by \SI{3.05}{\mega\hertz} to each selected pixel response in the previous step. The MATLAB fit function \verb+lsqcurvefit+ is used to perform a Levenburg-Marquadt non-linear fitting for each pixel ODMR response.
\item Histogram of distribution of resonant frequencies of individual pixels is analyzed. Pixels with artefacts or low ODMR response result in incorrect ODMR curve fits and have widely different resonant frequencies, as much as in gigahertz, from the median resonant frequency. On the contrary, all pixels with sufficient SNR levels have resonant frequencies clustered in a small 'continuous' band near the median resonant frequency. This resonant frequency range is visually inspected and provides bounds to the color axis of the resonant frequency maps. This simple bound removes pixels with wrong curve fits, pixels with low ODMR response and adjust dynamic range of the color axis of the 2D resonant frequency maps. 
\item The scaling of raw-data PL intensity data of the pixels to magnetic field time traces was done as described in the main text.
\item In temporal imaging datasets, pixels have an offset value ranging from 0 to 1024 (10-bit scale), but mostly centered  in the range of 500-600. Also, when a single resonant frequency is applied, heterogeneous pixels might show different baseline PL value. A small baseline time window of about 250-500\,\SI{}{\milli\second}, where no voltage was applied to the sample, was acquired and mean pixel value during the baseline window was subtracted from entire time trace of the pixel. Therefore, all pixels were centered to 0 in the beginning of temporal magnetic field tracking. Further, a 'detrend' in-built MATLAB function was applied over time traces of each pixel to remove linear drifts in the magnetic field tracking data.

\end{enumerate}

\subsection{Supplementary Videos}

The links to videos of imaging datasets in the main text have been provided below.  

\begin{enumerate}
\item Video1: Imaging video file of data shown in the main text Fig.\,\ref{fig:temporal1}. Scale bar \SI{27}{\micro\meter}. Microwire imaging, \SI{1.26}{\hertz}, 78 fps NV acquisition. \href{https://drive.google.com/file/d/1wA2RCD-kJXG-8HXN2C6DNiZXh57DAEYJ/view?usp=sharing}{Video link}.
\item Video2: Imaging video file of data shown in the main text Fig.\,\ref{fig:temporal2}. Scale bar \SI{34}{\micro\meter}. Microcoil imaging, \SI{17.9}{\hertz}, 78 fps NV acquisition. \href{https://drive.google.com/file/d/1avDm6U92DUX0EAOhLkYnanvw9msB-GrO/view?usp=sharing}{Video link}.
\item Video3: Imaging video file of data shown in the main text Fig.\,\ref{fig:APliketemporal}. Scale bar \SI{34}{\micro\meter}. Arbitrary waveform dynamics. In this video, each time frame was filtered with a MATLAB function 'filloutliers' to primarily reduce noise in the low-SNR pixels at the edges of the FOV. Outliers from image frames were removed on the criteria of 3 scaled median absolute deviation away from the median. Outliers pixels were filled with nearby local median value.  \href{https://drive.google.com/file/d/1bnS5pMr3x9ra7vzrUlrpwoej_8G5RRdp/view?usp=sharing}{Video link}.

\end{enumerate}

\subsection{Supplementary figures}
\begin{figure}[h]
\centering
\includegraphics[width=18cm]{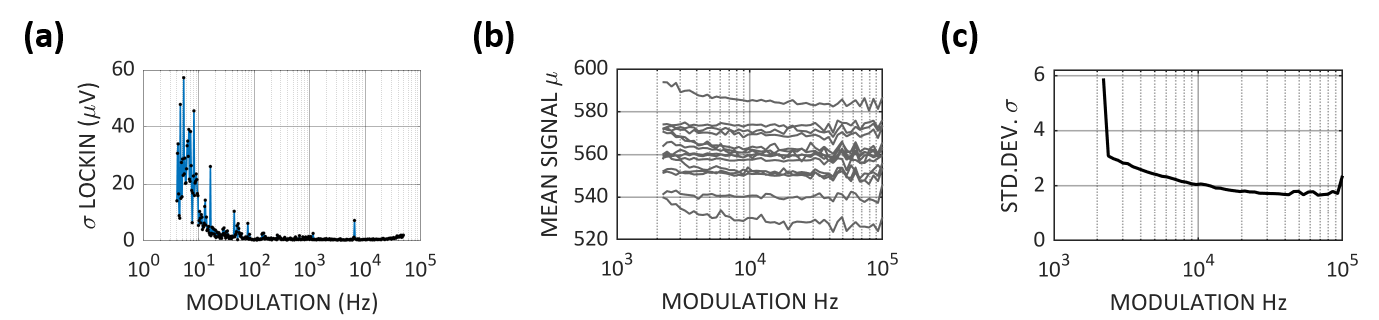}
\caption{Experimental noise spectrum of widefield imaging setup: (a) Noise spectrum of a reference single photodiode (SP) magnetometry setup across different diamond NV modulation frequency (which is same as lock-in amplifier reference). This spectrum has been measured with all experimental conditions identical to ODMR experiments except applied microwave excitation was off. Low-pass filter time constant set to \SI{100}{\milli\second} during the measurement. (b) Mean Noise measured for randomly chosen 15 pixels of the lock-in camera measured across different modulation frequencies. Similar to part A, experimental conditions were same as widefield ODMR experiments except microwave excitation was off. Units reflect 1024 (10-bit) points scale of camera output (c) Mean curve of standard deviation of randomly chosen 10000 pixels of the lock-in camera versus diamond NV modulation frequencies. Part (b) and Part (c) data obtained from same set of data of camera lock-in intensity frames ($n=20$) at different modulation frequencies ($n=20$ frames collected at each modulation frequency). Units reflected 1024 (10-bit) points scale of camera output. We note that the minimum camera lock-in frequency is \SI{2.2}{\kilo \hertz} , and therefore high noise at lower frequencies are not observed, unlike part (a) SP measurements.}

\label{fig:hc3_noisespectrum}
\end{figure}

\begin{figure}[h]
\centering
\includegraphics[width=18cm]{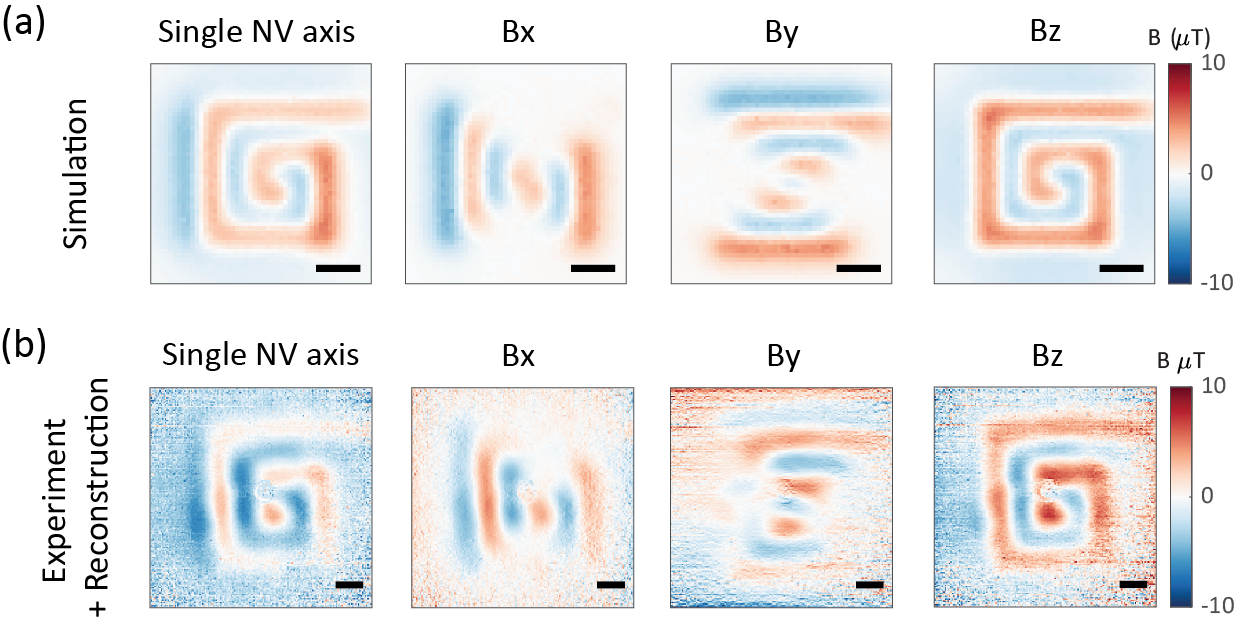}
\caption{Reconstruction of all three orthogonal axes \textbf{B} fields from single NV axis magnetic field image: (a) Simulated magnetic field profiles of the microcoil sample at sample standoff \SI{14}{\micro\meter}, current magnitude \SI{500}{\micro\ampere}. The simulated single NV axis projection image has been is shown on the same NV axis about which the widefield ODMR was acquired. Scale bar \SI{40}{\micro\meter}. (b) Experimentally obtained static magnetic field image of the microcoil current flow acquired about a single NV resonance peak with relatively higher magnetic field sensitivity. Orthogonal components of the magnetic field images reconstructed from the single NV axis magnetic image, assuming source free sensor plane and fourier inversion techniques. Scale bar \SI{34}{\micro\meter}   }

\label{fig:fourier_reconstruction}
\end{figure}

\begin{figure}[h]
\centering
\includegraphics[width=18cm]{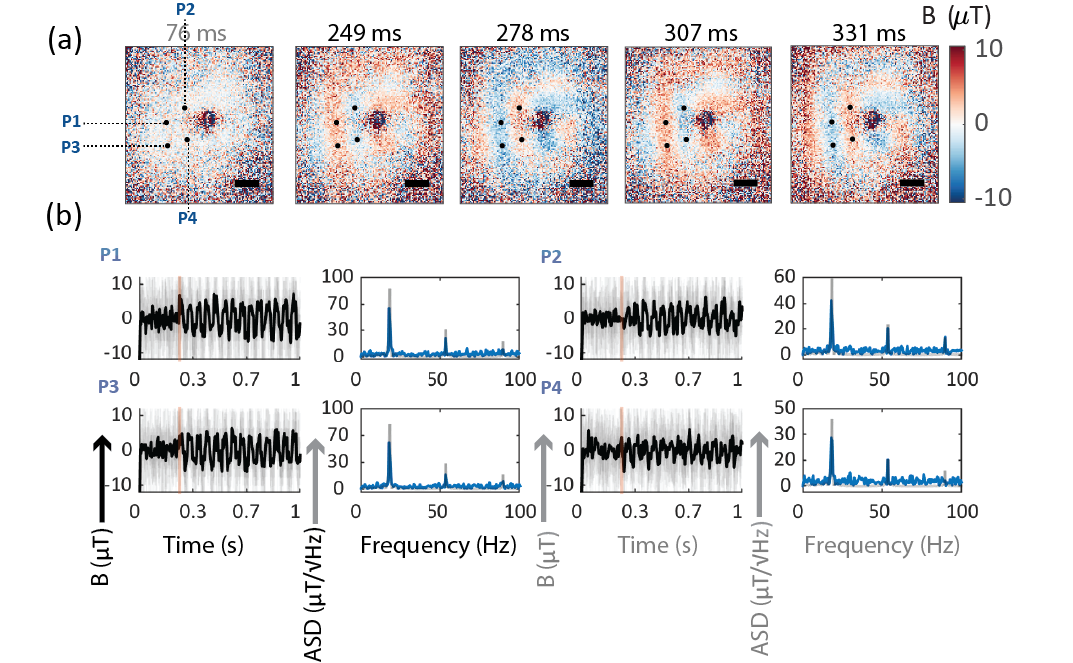}
\caption{Temporal imaging of \SI{18.9}{\hertz} magnetic field variation at 208 frames per second of the microcoil sample: (a) Magnetic field frames at single time-points (averaged $n=15$ iterations) showing alternating field image contrast with reversal in current direction. No voltage applied for a baseline time of \SI{0.25}{\second}, first frame selected from baseline window. A periodic square wave voltage waveform of alternating polarity was applied after the baseline time, at \SI{18.9}{\hertz} periodicity and peak current \SI{500}{\micro\ampere}. Exact magnetic field frame time-points have been shown on top of each image. Scale bar \SI{34}{\micro\meter} (b) Full temporal time traces of example pixels showing magnetic field tracking in time. Location of example pixels, labelled P1-P4 in the field of view have been shown in the magnetic field images. For each pixel, magnetic field traces versus time show tracking of applied the magnetic field, with faded gray lines as single iteration traces and solid black lines showing mean ($n=15$) magnetic field traces for the pixel. Amplitude spectral density of single-pixel field traces are shown on the left, where pixel Fourier spectra are in blue and applied voltage Fourier spectra has been shown in gray. Applied voltage spectral density is scaled to a constant to compare spectral content with pixel Fourier spectra. Since pixels track magnetic field, peaks in the pixel Fourier spectra matches with peaks in the Fourier spectrum of the applied voltage, with peaks occurring at magnetic field variation \SI{18.9}{Hz} and its odd harmonics. }
\label{fig:supple_uspiral_20Hz208fps}
\end{figure}

\begin{figure}[h]
\centering
\includegraphics[width=18cm]{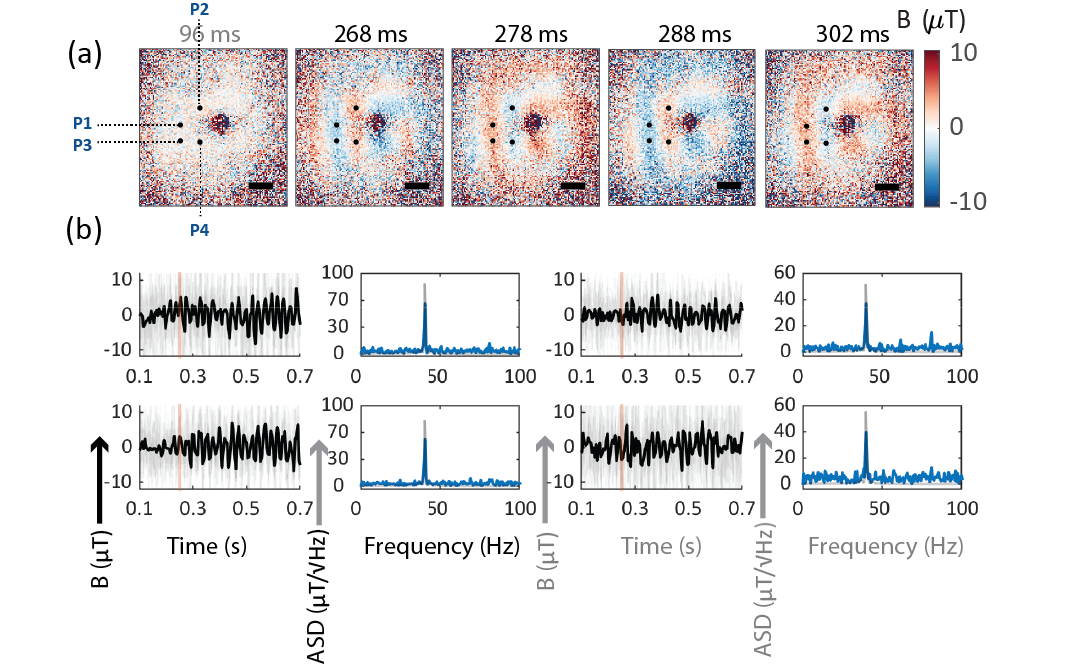}
\caption{Temporal imaging of \SI{41.52}{\hertz} magnetic field variation at 208 frames per second of the microcoil sample: (a) Magnetic field frames at single time-points (averaged $n=15$ iterations) showing alternating field image contrast with reversal in current direction. No voltage applied for a baseline time of \SI{0.25}{\second}, first frame selected from baseline window. A periodic square wave voltage waveform of alternating polarity was applied after the baseline time, at \SI{41.52}{\hertz} periodicity and peak current \SI{500}{\micro\ampere}. Exact magnetic field frame time-points have been shown on top of each image. Scale bar \SI{34}{\micro\meter} (b) Full temporal time traces of example pixels showing magnetic field tracking in time. Location of example pixels, labelled P1-P4 in the field of view have been shown in the magnetic field images. For each pixel, magnetic field traces versus time show tracking of applied the magnetic field, with faded gray lines as single iteration traces and solid black lines showing mean ($n=15$) magnetic field traces for the pixel. Amplitude spectral density of single-pixel field traces are shown on the left, where pixel Fourier spectra are in blue and applied voltage Fourier spectra has been shown in gray. Applied voltage spectral density is scaled to a constant to compare spectral content with pixel Fourier spectra. Since pixels track magnetic field, peaks in the pixel Fourier spectra matches with peaks in the Fourier spectrum of the applied voltage, with the peak occurring at magnetic field variation rate \SI{41.52}{Hz}.}

\label{fig:supple_uspiral_41Hz208fps}
\end{figure}

\begin{figure}[h]
\centering
\includegraphics[width=18cm]{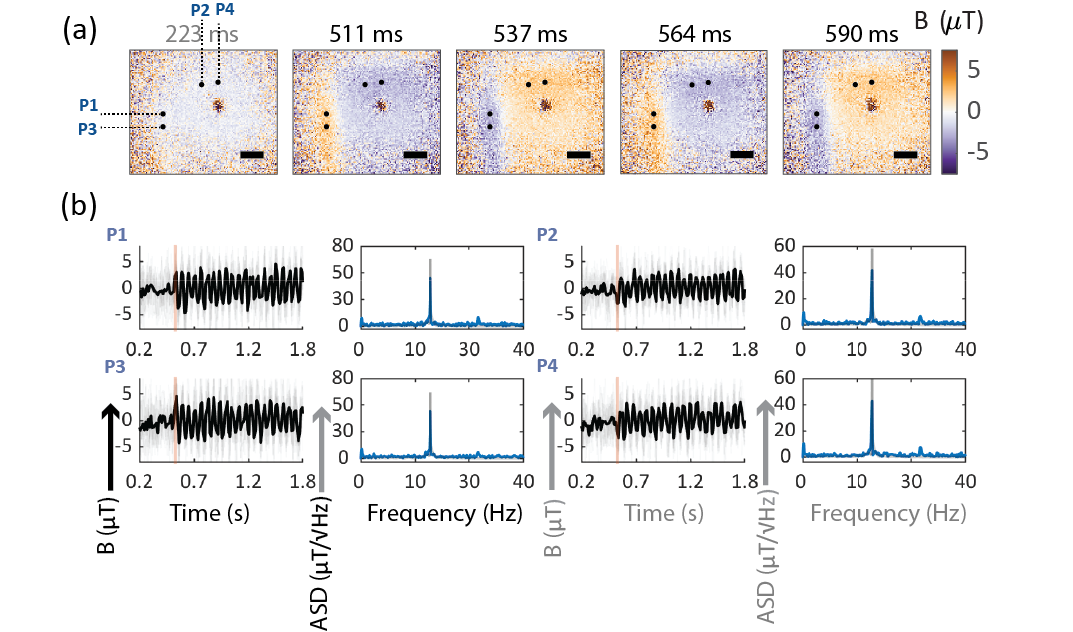}
\caption{Temporal imaging of \SI{16.3}{\hertz} magnetic field variation at 78 frames per second of the \SI{90}{\degree} bend microwire sample: (a) Magnetic field frames at single time-points (averaged $n=15$ iterations) showing alternating field image contrast with reversal in current direction. No voltage applied for a baseline time of \SI{0.5}{\second}, first frame selected from baseline window. A periodic square wave voltage waveform of alternating polarity was applied after the baseline time, at \SI{16.3}{\hertz} periodicity and peak current \SI{500}{\micro\ampere}. Exact magnetic field frame time-points have been shown on top of each image. Scale bar \SI{27}{\micro\meter}(b) Full temporal time traces of example pixels showing magnetic field tracking in time. Location of example pixels, labelled P1-P4 in the field of view have been shown in the magnetic field images. For each pixel, magnetic field traces versus time show tracking of applied the magnetic field, with faded gray lines as single iteration traces and solid black lines showing mean ($n=15$) magnetic field traces for the pixel. Amplitude spectral density of single-pixel field traces are shown on the left, where pixel Fourier spectra are in blue and applied voltage Fourier spectra has been shown in gray. Applied voltage spectral density is scaled to a constant to compare spectral content with pixel Fourier spectra. Since pixels track magnetic field, peaks in the pixel Fourier spectra matches with peaks in the Fourier spectrum of the applied voltage, with each pixel peak occurring at magnetic field variation \SI{16.3}{\hertz}. }
\label{fig:supple_uwire_29Hz78fps}
\end{figure}








\end{document}